\documentclass[prl,twocolumn,showkeys,superscriptaddress,footinbib]{revtex4}

\usepackage[T1]{fontenc}
\usepackage{amssymb}
\usepackage{amsmath}
\usepackage{xspace}
\usepackage{graphbox}
\usepackage{graphicx,xparse,booktabs}
\usepackage{ulem}

\usepackage[colorlinks=true,linktocpage=true,linkcolor=blue,citecolor=blue]{hyperref}
\usepackage{xurl}
\hypersetup{breaklinks=true}

\newcommand{\gpcyr}{$\rm{Gpc^{-3} \,yr^{-1}}$}

\begin{document}
\title{Binary neutron star mergers as the source of the highest energy cosmic rays}

\author{Glennys R. Farrar}
\email{gf25@nyu.edu}
\affiliation{Center for Cosmology and Particle Physics, Department of Physics \\ New York University, New York, NY 10003, USA}

\begin{abstract}

We propose that ultrahigh energy cosmic rays are produced in binary neutron star mergers.  This scenario can account for the heretofore inexplicable narrow rigidity range of UHECRs, because the jets of BNS mergers are generated by a gravitationally-driven dynamo and thus are nearly identical due to the narrow range of BNS masses.  Observed UHECRs with energies well beyond 100 EeV can be explained as $r$-process nuclei, without invoking an exotic source class.  Evidence for this mechanism, and its prediction of coincidences between neutrinos above 10 PeV and gravitational waves, are discussed.

\end{abstract}

\date{\today .~Phys. Rev. Lett. in press; submitted May 20, 2024; accepted Jan. 21, 2025}

\maketitle

{\bf \noindent Introduction}\\ \label{sec:intro}
Major advances in ultrahigh energy cosmic ray (UHECR) observations in the past two decades have lead to a much better understanding of their spectrum and composition, but the central question remains:  What are the systems responsible for accelerating cosmic rays to the highest observed energies?   
We argue here that two particularly puzzling features of the observations strongly suggest that extragalactic cosmic rays are produced in binary neutron star (BNS) mergers, with the very highest energy events originating as $r$-process nuclei such as tellurium.  This BNS-merger scenario explains all current observations, within present uncertainties. We outline tests of the scenario.

Options for UHECR sources are constrained by the observed rate and energetics, spectrum and composition, inhomogeneities in arrival directions (dipole and higher multipoles and hotspots), and by individual extremely high energy events such as the recent Amaterasu particle~\cite{TAamaterasu} and earlier Fly's Eye "OMG" event~\cite{FlysEye95}.  Crucially, UHECR accelerators have an essentially universal maximum rigidity, $\mathcal{R}\equiv E/Ze$, where $E$ is energy and $Ze$ is charge.  Furthermore, under the conventional assumption that the heaviest UHECRs are iron nuclei, an entirely different source class is required to explain the very highest energy events.  After giving a brief overview, we review the constraints on UHECR sources and note why previously proposed source candidates are problematic in one way or another, motivating the need to seek new possibilities.  We show that BNS-mergers can explain all of the observations as of now.  We close with a discussion of tests and implications of this scenario.

{\bf \noindent Overview}\\
The very large exposures of the Pierre Auger Observatory in Argentina and the Telescope Array in Utah have resulted in $\approx$ 50,000 events having been recorded above 10 EeV ($10^{19}$ eV)~\cite{GollupAD_ICRC252}, of which more than 40 are above 100 EeV~\cite{augerTop100}.  Above 10 EeV, Galactic cosmic rays  (GCRs) make a negligible contribution.  
The cutoff in the spectrum at the highest energies is due to a combination of limited accelerating-power of the sources and Greisen-Zatsepin-Kuzmin processing~\cite{Greisen66,ZK66,AugerCombinedFit23}.  For protons, GZK energy loss is due primarily to inelastic scattering on cosmic microwave background photons.  For nuclei, interactions with the cosmic microwave background (CMB) and other extragalactic background light cause spallation or breakup; these interactions conserve the energy per nucleon. For common nuclei other than protons, the rigidity, $\mathcal{R}\equiv E/Ze$, is twice the energy per nucleon and is thus approximately conserved during propagation.

The observed composition above 10 EeV is consistent with UHECRs being nuclei with a narrow range of rigidities centered on $\mathcal{R}\approx 4$ EV; see Fig.~\ref{fig:rig_spec} in Sec. 1 of~\cite{SupMat}.
The light extragalactic component below 10 EeV is well-explained as resulting from spallation of primary UHE nuclei as they pass through the environment surrounding the accelerator~\cite{ufa15}.  

{\bf \noindent Constraints on UHECR sources}\\
\noindent \uline{1.  Universal maximum rigidity:}
\label{sec:constraints}
The most puzzling feature of UHECRs is that the range of nuclear masses present at any given energy is small~\cite{augerComposition14}, so the rigidity spread is very narrow -- only a factor-few in width~\cite{bf23}.  Uncertainties in modeling hadronic interactions in the relevant forward kinematic regime mean that the absolute value of the central mass at any given energy is uncertain at up to the $\approx 20$\% level. However the fractional {\it spread} in mass is largely model independent and is unambiguously narrow (Sec. 1 of~\cite{SupMat}).

Different nuclei accelerated in an individual source should have essentially the same spectrum in rigidity, since known acceleration mechanisms depend only on rigidity and that is approximately preserved during propagation as discussed above.  However for the source types proposed up to now, different sources should exhibit different maximum rigidities so that an ensemble of sources would in general have a broad rigidity spectrum, contrary to observations~\cite{AugerCombinedFit23}.  
This problem was systematically investigated in Ref.~\cite{foteini+23}, where it was shown that none of the conventional source candidates [blazars, long gamma ray bursts (GRBs), tidal disruption events (TDEs)] have a  sufficiently small population variance to yield a distribution of maximum rigidities consistent with the UHECR composition data.  [Seyfert galaxies -- low luminosity active galactic nuclei (AGN) -- were found to be marginally compatible within the scope of the analysis but likely excluded when all sources of variance are taken into account~\cite{foteini+23}.] 

\label{sec:reqts}
\noindent \uline{2. The Hillas criterion:} A candidate UHECR source must be capable of accelerating UHECRs to the observed rigidity.  The characteristic size of the accelerating region, $R$, must be large enough to confine the UHECR -- i.e., $R$ cannot be significantly smaller than the Larmor radius of the cosmic ray.  This leads to the {\it Hillas criterion} relating the maximum rigidity to which UHECRs can be accelerated, to the size and field strength in the source: 
\begin{equation}
\label{eq:hillas}
   \mathcal{R}_{\rm max,EV} \lesssim  3 \times 10^{-16}\, \Gamma\,R_{\rm cm} \,B_{\rm G} ~, 
\end{equation}  
where  $B_{\rm G}$ is the magnetic field strength in Gauss, $\Gamma$ is the relativistic boost factor and $R_{\rm cm}$ is the characteristic size of the confinement region in cm; see, e.g.,~\cite{fg08}.  Since the Poynting luminosity of a jet is proportional to the cross-sectional area times $\vec{E}\times \vec{B}$, the Hillas 
condition~\eqref{eq:hillas} on $R  B$ means that for jet acceleration, the minimum Poynting luminosity is determined, independently of $R$ and $B$ separately.  This implies the condition on the (isotropic equivalent) total power of a jet capable of accelerating CRs to a rigidity $\mathcal{R}_{\rm max, EV}$~\cite{fg08}: 
\begin{equation}
\label{eq:Lbol}
    \mathcal{L}_{\rm bol} > \frac{c}{6} \,\Gamma_{\rm jet}^4B^2R^2 \approx 10^{41} \,\Gamma_{\rm jet}^2 \,\mathcal{R}_{\rm max, EV}^2\, {\rm erg/s}~.
\end{equation}
(If UHECRs were protons so $\mathcal{R}= E$, this would be a very stringent requirement and rule out all but extremely luminous sources.)   The locus of UHECR production within a source region satisfying Eqs.~\eqref{eq:hillas} and ~\eqref{eq:Lbol} is further constrained by the requirement that the field strength is low enough that synchrotron loss is subdominant to energy increase from acceleration~\cite{greisenICRC65}.  

\noindent \uline{3. Total UHECR energy injection:} A source class capable of explaining all UHECRs must account for the observed total flux of UHECRs. Since UHECRs lose energy during propagation due to GZK processes, a steady replenishment is required.  The volumetric energy-injection rate for UHECRs above 10 EeV is~\cite{mfu22}
\begin{equation}
\label{eq:Qdot}
\dot{\mathcal{Q}} \approx 6 \times 10^{44} \text{ erg\, Mpc}^{-3} \,\text{yr}^{-1}~.
\end{equation}
(The low energy limit chosen to define $\dot{\mathcal{Q}}$ only matters at the $\mathcal{O}(1)$ level; 
the 10 EeV threshold adopted here avoids contamination by Galactic cosmic rays.)  For continuous sources, Eq.~\eqref{eq:Qdot} 
 constrains the product of the number density of sources times their mean luminosity in UHECRs:  $n_S \, \mathcal{L}_{\rm UCR} = \dot{\mathcal{Q}}$.  For transient sources, the constraint is $\Gamma_S \, \mathcal{E}_{\rm UCR} = \dot{\mathcal{Q}}$, where $\Gamma_S$ is the volumetric rate of the transient sources and $\mathcal{E}_{\rm UCR}$ is the total energy in UHECRs above the specified threshold, produced in an average transient event.  
   
\noindent \uline{4. Effective number density of contributing sources:}   
Auger finds that the higher multipoles beyond dipole are consistent with an isotropic distribution; this upper limit on small and intermediate scale inhomogeneities was used in Ref.~\cite{bf23} to constrain the density of UHECR sources to 
  \begin{equation}
  \label{eq:nsrc}
 n_S \gtrsim 10^{-3.5}\,{\rm Mpc}^{-3}.
 \end{equation}
This number density constraint disfavors rare continuous sources such as blazars, powerful AGN and starburst galaxies, apart from a marginally viable narrow window for sufficiently high extragalactic magnetic smearing~\cite{bf23}.   
 
For transients, the constraint~\eqref{eq:nsrc} is on the effective source density $n_S^{\rm eff} \equiv \Gamma_S \,\tau$, where $\Gamma_S$ is the volumetric rate introduced above and $\tau$ is the mean arrival time spread of the UHECRs from a transient event within the GZK horizon; the superscript ``eff" is suppressed below~\cite{JettedSrc}.
Because of their deflections in extragalactic magnetic fields, UHECRs arriving at the Milky Way are spread over an angle $\psi$ and arrive with time delay up to $\tau$~\cite{Achterberg:1999vr}:
\begin{align}
\label{eq:psi}
\psi &\approx 30^\circ \beta_{\text{EGMF}}\sqrt{D_{\textrm Mpc}}/\mathcal{R}_{\rm EV}  \\
    \label{eq:taueff}
    \tau &\approx 0.14 \, (D_{\textrm Mpc} \beta_{\text{EGMF}}/\mathcal{R}_{\rm EV})^2 \,{\textrm{Myr}}~,
\end{align}
where $\beta_\mathrm{EGMF} \equiv B_\mathrm{EGMF} / \mathrm{nG}\,\sqrt{L_c / \mathrm{Mpc}}~
$ is the extragalactic magnetic smearing parameter.
Commonly adopted estimates are $B_\mathrm{EGMF}\approx 1 $ nG, and $L_c$ in the range 0.01-1 Mpc, giving $\beta_\mathrm{EGMF}\approx 0.1 - 1$.
Long GRBs and visibly jetted TDEs 
are disfavored as the exclusive source of UHECRs by this criterion, being marginally possible only if the mean $\beta_{\text{EGMF}}$ is near its maximum allowed value~\cite{bf23}. 

\noindent \uline{5.  Amaterasu Event:} 
The very energetic cosmic ray Amaterasu detected by the Telescope Array~\cite{TA_Amaterasu} provides direct evidence that at least some UHECRs are produced by transient events. GZK energy losses during propagation severely restrict the distance of Amaterasu's source. 
Taking into account the uncertainties in energy and composition, Ref.~\cite{ufAma23} backtracked Amaterasu through the Galactic magnetic field (GMF) using the recent UF23 suite of GMF models~\cite{UF23}, designed to span the uncertainty in GMF deflections.  The maximum source volume contains only ordinary galaxies and a few low-luminosity AGNs, even conservatively extending the horizon to 125 Mpc motivated by its possible identification as an $r$-process nucleus as discussed below. 

{\bf \noindent Source candidates}\\
\label{sec:ConvCands}
The jets found in powerful AGN or long GRBs~\cite{Vietri_GRB-UHECR_95,waxman95}, early favorite UHECR sources, are now constrained to be minor contributors at best, based on the source density requirement, $ n_S \gtrsim 10^{-3.5}\,{\rm Mpc}^{-3}$, their absence in the Amaterasu source volume~\cite{ufAma23}, and the requirement that individual sources have a common peak rigidity~\cite{foteini+23}.  Additionally, long GRBs can only account for the volumetric energy injection rate, $\dot{\mathcal{Q}}$, if the energy in UHECRs exceeds the energy in gamma rays by two orders of magnitude~\cite{fg08}. 

Low-luminosity AGN, minimally satisfying the bolometric luminosity condition to accelerate UHECRs to 5 EV with $\mathcal{L}_{\rm bol}  \gtrsim 10^{42.5}$ erg/s, may be marginally consistent with the number density requirement -- c.f., the AGN luminosity function in~\cite{duras+AGN20}.  But 
lower luminosity AGN are even more abundant (Zaw et al.~\cite{zcf19} finds that 40\% of the SDSS galaxies out to $z = 0.07$ are either broad-line AGN or satisfy the Kewley narrow-line AGN criterion) so the narrow observed rigidity distribution of UHECRs is problematic for the LLAGN scenario;  see also~\cite{foteini+23}.     

Another possibility for UHECR acceleration, suggested in~\cite{fg08}, are short-lived but powerful jets due to tidal disruption events (TDEs) when a star passes close enough to a supermassive black hole that it is shredded and forms an accretion disk and jet, or giant flares of  AGNs produced by a disk instability or accretion event. TDEs were subsequently discovered~\cite{vVelzen+TDE11}, the rate found to be potentially sufficient~\cite{vVfTDErate14}, and many TDEs have now been detected.  Several spatio-temporal correlations between TDEs and astrophysical neutrinos have been reported~\cite{stein+TDENature20,Reusch+PRL22,Jiang+TDEnu23}.  Jetted TDEs appear to be too rare to satisfy the energy injection and source density requirements [Eqs. ~\eqref{eq:Qdot} and 
~\eqref{eq:nsrc}], but late-emerging jets may resolve that; see~\cite{piran+UCR23} for a recent discussion.  However  the near-universal rigidity of UHECRs excludes TDEs being a significant source of UHECRs because TDEs comprise a very diverse population differing enormously from one event to another in the properties determining the character of a jet:  primarily, the mass, radius and pericenter distance of the disrupted star.  Indeed, most TDEs do not even produce jets, let alone standard ones. 

UHECRs might be accelerated in accretion shocks of galaxy clusters and filaments~\cite{Simeon+ICRC23}, but Amaterasu originated in a region with no galaxy cluster; therefore mild as well as massive shocks must accelerate to the highest energies, which implies a very broad distribution of source maximum rigidities, contrary to observation.

{\bf \noindent Binary Neutron Star mergers and UHECRs}\\
\label{sec:BNSmergers}
Binary neutron star mergers provide a natural rationale for the near-universal rigidity of UHECRs.  The merger of two 1.35 $M_\odot$ neutron stars creates a pair of powerful back-to-back jets easily satisfying the Hillas criterion and Poynting luminosity requirements. The ultra-high-resolution neutrino-radiation-GR-MHD simulation of Ref.~\cite{Kiuchi+BNSdynamo23} demonstrated that these magnetized jets are produced by a gravitationally-driven dynamo.  Therefore, given that the range of total mass of measured double NS systems is very narrow, with just a 5\% spread around the mean~\cite{freireFootnote}, it is plausible (but not assured) that 
the jets and outflows produced in their mergers are nearly identical from one to another, at least as far as UHECR-acceleration is concerned~\cite{sGRBfootnote}.  
The gravitational wave signal has been shown to have little sensitivity to the initial NS spin configuration~\cite{East+GWSpinDep19}, but simulations like~\cite{Kiuchi+BNSdynamo23} for different initial NS spins are needed to quantify the expected variation in jet and magnetized outflow properties within the population and to determine whether NS-BH mergers may also produce capable jets.   

Both the effective source density and the UHECR energy injection rate can be satisfied by BNS mergers.  They both  depend on the BNS merger rate.  Two different methods have been used to estimate the merger rate, giving similar results albeit presently with large uncertainties.  The LIGO-Virgo-Kagra collaboration's conservative range on the BNS merger rate is 10-1700 \gpcyr~\cite{LVK_binaryMergerRate23},  and the recent detailed study of the afterglows of 29 short GRBs~\cite{SGRBrate23} finds, using events with well-measured opening angles, a beaming-corrected short GRB rate of 
$1786^{+6346}_{-1507}$ \gpcyr.  Below, we take the BNS merger rate to be $\Gamma_{\rm BNS} = \Gamma_3 \,
\! \times \! 10^3 \,{\rm Gpc}^{-3}\, {\rm yr}^{-1}$.

The observed source density requirement is easily satisfied, given that the expected range for the magnetic smearing parameter is $\beta_\mathrm{EGMF}\approx 0.1 - 1$ and $ \beta_{\rm EGMF}  \gtrsim 0.04$ is sufficient even for the minimum BNS merger rate.  

The observed UHECR energy injection rate is marginally compatible with present constraints.  Reference~\cite{Kiuchi+BNSdynamo23} finds a Poynting luminosity $>\! 10^{51}$ erg/s and a magnetically-driven wind with kinetic energy $5\!\times \! 10^{50}$ erg, when the simulation ends 150 ms post-merger.  For a one second duration jet, with $\epsilon_{\rm CR}$ conversion efficiency to UHECRs, the observed $\dot{\mathcal{Q}}$ is reproduced for $\epsilon_{\rm CR} \Gamma_3 = 0.6$;  see also~\cite{MuraseFukugita18}.  Simulations of particle acceleration including feedback on the medium, extending~\cite{ComissoSironi19,ZhangSironiGiannios21,Li+Accel23}, will clarify the efficiency of jet acceleration.  

Another scenario to consider is UHECR production in the longer-lasting, highly magnetized outflow {\it outside} the jet.  
As recently shown, acceleration in magnetized turbulence produces a sharply cutoff, $E^{-2.1} {\rm sech} [ (E/E_{\rm cut})^2 ] $ spectrum~\cite{cfm24}, which gives a much better fit to the Auger spectrum and composition than obtained with the exponential cutoff characterizing diffusive shock acceleration~\cite{protheroeStanev99}.  Acceleration in magnetized turbulence is maximally efficient, with an acceleration timescale $\tau_{\rm acc} \approx 1.6 \, l_{\rm coh} /c$, independent of energy; moreover half the dissipated magnetic energy is transferred to cosmic rays~\cite{cfm24}.   
After a BNS merger, the field strength is initially large and the ion energies stay small, limited by synchrotron emission.  As the outflow expands the field strength drops and the coherence length increases;  the characteristic maximum ion rigidity for a given charge and mass, which is set by the balance between synchrotron emission and acceleration rates, grows as the system expands.  After the field drops below a critical value, depending on rigidity and composition, synchrotron losses are subdominant to acceleration and the rigidity cutoff is determined by the limitation of the accelerator: $\mathcal{R}_{\rm cut} \approx 0.65 \, B\, l_{\rm coh}$~\cite{cfm24}.  A quantitative estimate for the BNS merger case, normalized based on the simulation of~\cite{Kiuchi+BNSdynamo23}, gives encouraging agreement with the observed value of $\mathcal{R}_{\rm cut}$ (GRF, paper to appear shortly).   Particle-in-cell simulations including synchrotron emission and modeling the expansion of the magnetized plasma are needed for a more definitive comparison to data and to estimate the total energy in UHECRs.
 
\label{sec:tests}
{\bf \noindent Tests and implications}\\
\noindent \uline{ $R$-process UHECRs and the highest energy CRs:}  In the BNS-merger scenario,  UHECRs are a combination of nuclei synthesized in the merger and typical nuclei of the ambient medium around the binary, which get swept up and accelerated in the jet or magnetized turbulent outflow; these UHECRs are then spallated to some degree while escaping through the medium surrounding the accelerator and during extragalactic propagation~\cite{ufa15}.  About half of all $r$-process nuclei are produced by nucleosynthesis in binary neutron star mergers in a fat equatorial disk~\cite{Kasen+R-processBNS17}, potentially greatly enhancing the fraction of $r$-process elements in UHECRs relative to their Galactic average abundance.  The masses of $r$-process nuclei peak around $A =130$ (Te-Xe; Z = 52-54) and $A=195$ (Pt; Z = 78); see~\cite{Cowan+r-processRMP21} for a recent review.   Taking $\mathcal{R}\approx 4.5\pm 0.5$ EV to approximate the mean universal rigidity~\cite{bf23}, the energies of these peak $r$-process nuclei after acceleration in the source are $E \approx 240 \pm 25$ EeV and $350\pm 40$ EeV.

The proposal here that UHECRs originate in BNS mergers, and hence naturally contain some component of $r$-process elements, provides the first compelling, non-exotic explanation for the highest energy UHECRs ever recorded: Amaterasu at 210-250 EeV~\cite{TA_Amaterasu,ufAma23} and the Fly's Eye ``OMG'' event~\cite{FlysEye95} at 250$\pm 70$ EeV using the modern air fluorescence yield~\cite{AIRFLY}. Since the rigidity range of observed UHECRs is very narrow, we can infer the charge of these events for any given energy using the mean rigidity of 4.5 EV~\cite{bf23}.  This results in an estimated charge of $(53\pm 15)  \left(\frac{4.5}{\mathcal{R}}\right)$, suggesting these events originated near the Te-Xe $r$-process peak, since the charge loss of heavy nuclei in propagation is small~\cite{AmaFoot}.  
The depth-of-shower-maximum ($X_{\rm max}$) of Amaterasu is not known,  but  $X_{\rm max}$ of the OMG event ($ 815^{+60}_{-53}\, {\rm g\, cm}^{-2}$ including  systematic and statistical uncertainties \cite{FlysEye95}) is consistent with a Te interpretation:  $ \langle X_{\rm max}\rangle$ shifts with nuclear mass and incident cosmic ray energy as $d \langle X_{\rm max}\rangle /d \,{\rm log_{10} A} \approx - d \langle X_{\rm max}\rangle /d \,{\rm log_{10} E} \approx - 60 \,{\rm g\,cm}^{-2}$~\cite{kampertUnger12}; using Sybill23d to simulate the shower development for Fe at a reference energy, predicts $X_{\rm max} = 780 \, {\rm g\, cm}^{-2}$ for Te at 250 EeV.  
  
By contrast, if the most massive UHECRs are Fe rather than $r$-process, the OMG and Amaterasu events would have rigidity $\mathcal{R} \gtrsim 10$ EV -- more than twice that of other UHECRs -- and an unknown exotic origin~\cite{ES95,Risse+FE04,Gnatyk+FEinGMF16,Fitoussi+FE19}.  By Eq.~\eqref{eq:Qdot}, their sources would require at least an order of magnitude higher luminosity than other events.   In short, the very existence of the observed OMG and Amaterasu events is evidence for the binary neutron star merger origin of UHECRs.  With a large enough sample of well-measured events above 150 EeV, as proposed with future detectors~\cite{GCOShorandel21}, the contributions of different $r$-process nuclei to extreme energy CRs could be mapped out. 

\noindent  \uline{Extremely High Energy neutrinos associated with BNS mergers:}   As discussed, the light composition of UHECRs below the ankle is nicely explained by spallation of higher-mass UHECRs in the environment of the accelerator~\cite{ufa15}. These interactions produce a population of EHE neutrinos with peak energy of about 20 PeV~\cite{mfu22}.  Thus in our BNS merger scenario, every neutrino detected in the 20 PeV region should be preceded by a gravitational wave (GW) produced by the neutron star merger (absent some not-yet-identified additional source of EHE neutrinos).  Since the UHECR, GW and neutrino travel at essentially the speed of light, the delay between GW and neutrino arrivals arises from the time required for the CR to become ultrarelativistic and the extra path length of the cosmic ray due to magnetic deflections prior to its neutrino-producing interaction in the source environment.  
No neutrinos were detected in coincidence with GW170817~\cite{BNS17}, consistent with expectations.  Happily, coincident observations of a BNS merger and a neutrino in this very high energy population, which would confirm the BNS origin of UHECRs, look promising with next-generation detectors;  see Sec. 3 of~\cite{SupMat} for details.

{\bf \noindent Summary}\\  
\label{sec:Sum} 
We have proposed that binary neutron star mergers are the source of all or most ultrahigh energy cosmic rays.  The very narrow range of total mass of binary neutron stars, and the fact that the post-merger magnetic fields are produced by a dynamo whose energy source is gravitational, provides the first rationale for why UHECR rigidities can be almost universal across different sources -- a severe challenge to previously proposed source types, as demonstrated in~\cite{foteini+23}. From the observed narrow range of UHECR rigidities we can deduce that previously considered source types, which have a wide range of maximum rigidities, make at most a subdominant contribution to UHECR production. Present information on the rate of binary neutron star mergers and their capacity to produce UHECRs is compatible with all extragalactic cosmic rays originating in BNS mergers. 

Since neutron star mergers are the primary site of $r$-process nucleosynthesis, our proposal suggests the existence of UHECRs with a composition significantly heavier than iron.  An immediate success of this scenario is that the Fly's Eye ``OMG'' and Amaterasu events are naturally explained as originating from  $r$-process nuclei in the tellurium peak, rather than being extreme outliers requiring a uniquely powerful accelerator or exotic origin.  In the future, UHECR data should be interpreted allowing for a component of $r$-process nuclei, and codes to model UHECR energy losses during extragalactic propagation should be extended to include $r$-process nuclei.

This Letter shows that UHECR production in BNS mergers  satisfies global observational requirements.  The mechanism of UHECR acceleration within the merger ejecta, and an attempt to estimate the fraction of $r$-process nuclei, will be the topic of future works.     

\section*{Acknowledgments}
I am grateful to T. Bister, M. Muzio, M. Unger and L. Comisso for fruitful collaborations and for valuable input which has stimulated and enriched the work presented here. I also thank B. Dawson,  M. Unger and A.~A.  Watson for discussions of an updated estimation of the energy of the Fly's Eye OMG event, helpful information from M. Mapelli, A. MacFadyen, B. Metzger, E. Most and K. Murase, and useful suggestions from the referees which have improved the clarity of the paper.
This research has been supported by National Science Foundation grants PHY-2013199 and PHY-2413153 and performed in part at the Aspen Center for Physics,  supported by NSF PHY-2210452.


\begin{thebibliography}{64}
\bibitem{JettedSrc} For jetted sources, continuous or transient, the effective contributing source density is further reduced relative to the actual density by the fraction $f_{\rm contrib}$, which may be crudely estimated as $f_{\rm contrib} \approx {\rm sin}( \sqrt{\delta_j^2 + \psi^2})$ with $\delta_j$ being the jet opening angle.  The mean source distance of UHECRs above 32 EeV is 70 Mpc~\cite{bf23}, so for a 2-sided jet whose opening angle is 10$^\circ$ the sky-averaged value of $f_{\rm contrib}$ ranges from 0.2 for small $\beta_{\text{EGMF}}$  to 1 for $\beta_{\text{EGMF}}\gtrsim 0.6$
 \BibitemShut {NoStop}%
\bibitem{sGRBfootnote} The wide variation in the light curves of short GRBs does not contradict this; simulations constrained to the kilonova associated with the binary neutron star merger GW170817 show that the variation in the observed light curves of short GRBs is attributable to different viewing angles of the jet and different circumbinary media~\cite{MacF+Jets18}
 \BibitemShut {NoStop}%
\end{thebibliography}

%



\clearpage
\onecolumngrid
\begin{center}
Supplementary Materials\\ 
\vspace{0.08in}
{\large \bf{Binary Neutron Star Mergers as the Source of the Highest Energy Cosmic Rays} }\\
\bigskip
Glennys R. Farrar\\
Center for Cosmology and Particle Physics, Department of Physics \\ New York University, New York, NY 10003, USA
\end{center}

\bigskip

\twocolumngrid

\section{1.  Narrow rigidity Distribution}

The nuclear mass, $A$, of a UHECR can be constrained from the longitudinal development of the air shower it produces in the atmosphere.  A heavy nucleus interacts quickly and breaks up, producing many sub-showers.  Thus there is little event-to-event variation between showers if all the nuclei are heavy.  By contrast, protons can interact quickly or penetrate deeply before their first interaction, and the interaction may be diffractive with little particle production and energy loss, or central with many secondaries.  These facts of proton interactions leads protons to have on average deeper-developing showers and much greater shower-to-shower variance in $X_{\rm max}$ compared to the air showers of heavy nuclei.  Hence the combination of the mean depth-of-shower maximum, $<\!X_{\rm max}\!>$, and its variance, $\sigma(X_{\rm max})$, reflect the mean mass of primaries as well as the spread of masses at a given energy.   See the review~\cite{kampertUnger12} for additional insights on interpreting air shower data.

\begin{figure}[h]
\hspace{-3.1in}
 \includegraphics[trim={0.8in 0 0 1.2in},clip,width=0.45\textwidth]{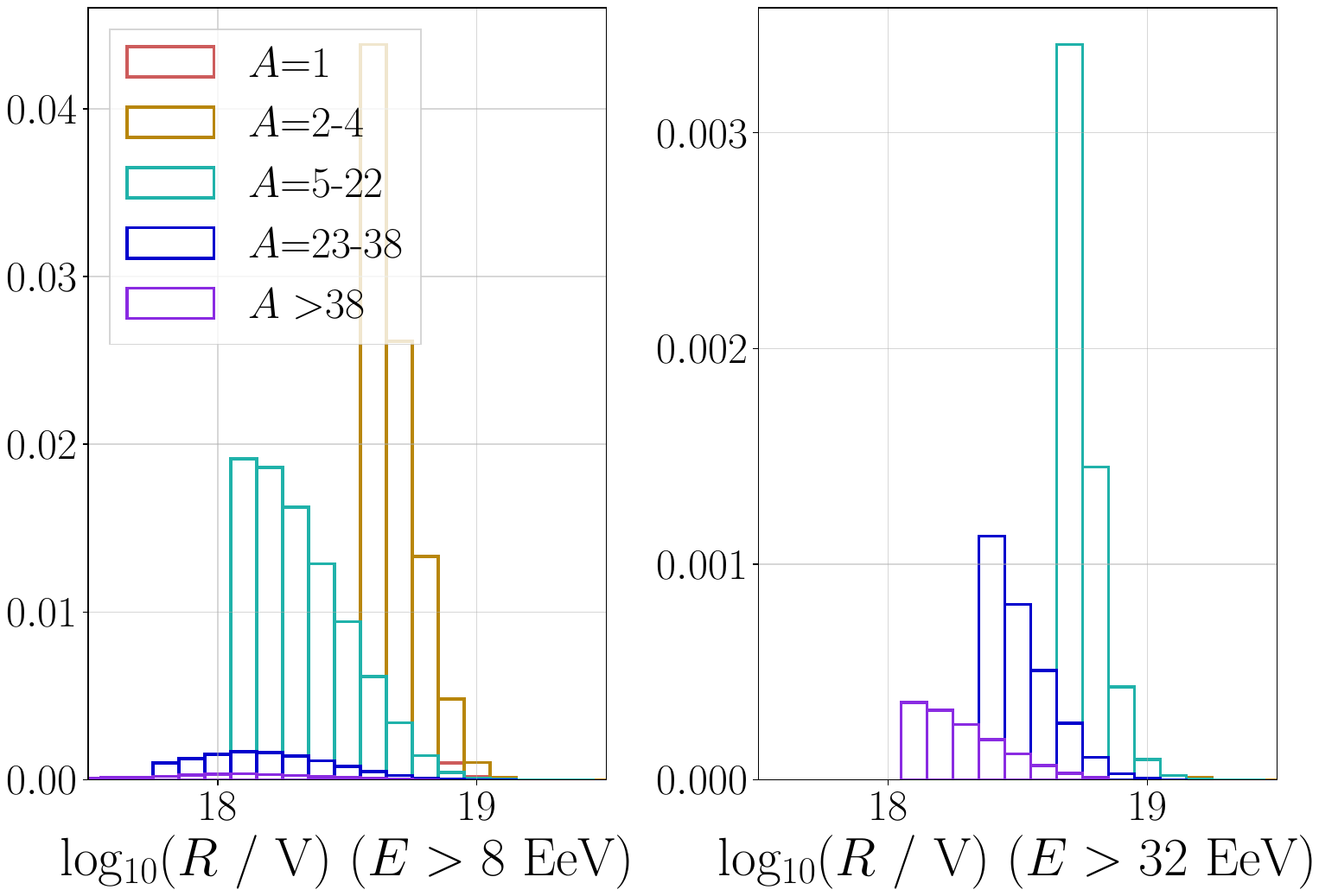}
    \caption{Fig.~4 of Ref.~\cite{bf23} showing the frequency (in arbitrary units) of arriving rigidities on Earth, for two energy thresholds, $E>8$\,EeV (\textit{left}) and $E>32$\,EeV (\textit{right}); the mean rigidities are 3.5 and 4.5 EV for the two datasets, respectively. The multi-peak structure is due to the coarse binning in mass following Auger.} 
    \label{fig:rig_spec}
\end{figure}

Mapping $<\!X_{\rm max}\!>$ onto mean composition has uncertainties from the hadronic interaction modeling, but whether the variance is small or not is a robust,  model-independent indicator of the range of masses contributing at a given energy.  The data are now very precise, and show that only a narrow range of masses, and hence of rigidities, is present at any given energy. Furthermore, the mean rigidity and its spread, of order $5\pm 3$ EV, depends only very weakly on energy bin.  This can be seen in Fig.~\ref{fig:rig_spec}, reproduced from Ref.~\cite{bf23}. Subsequent to~\cite{bf23}, Auger's analysis using a Deep Neural Network to extract $X_{\rm max}$ and $\sigma(X_{\rm max})$ from the 10-times-higher statistics Surface Detector data has appeared; it further strengthens the evidence for a narrow spread in rigidities~\cite{augerSDcompo24}. Note that the DNN was trained on simulated showers including only $A\leq 56$ and hence (like all experimental composition analyses up to now) the results of ~\cite{augerSDcompo24} cannot be used at face-value to assess the composition at the highest energies if, as proposed here, nuclei beyond Fe may be present.  Future re-analyses may be able to discriminate whether data favors a continuing increase in the mean mass or termination with Fe.

\section{2.  Neutrino-Gravitational Wave Coincidences}

As discussed in the main text, a strong prediction of the proposal that all UHECRs are produced in BNS mergers, is that every neutrino in the 20 PeV peak produced by UHECR interactions has an associated gravitational wave.  
In this section we derive an upper limit on the fluence of EHE neutrinos accompanying GW170817, produced by UHECR interactions as they leave the source environment, and compare to the multimessenger limits on neutrinos accompanying GW170817 from ANTARES, IceCube, Auger and LIGO~\cite{Auger+GWNeutrinos17}.  We also search for short GRB coincidences with observed neutrinos above 5 PeV, and take a first look at prospects for coincidences with future neutrino and GW detectors.

The time delay between arrival of a GW signal and the neutrino depends on the time scale for accelerating the CR to ultrarelativistic velocity and the increased  CR path length due to magnetic deflections before the interaction which produces the pion.  Since these are presently uncertain and could be large, in the following we place no temporal association requirement apart from the GW or sGRB preceeding the neutrino event.  Future measurement of the time delay spectrum will give important information on the UHECR production locus and environment.

\begin{figure*}[!]
\hspace{-17cm}
    \includegraphics[width=0.9 \textwidth]{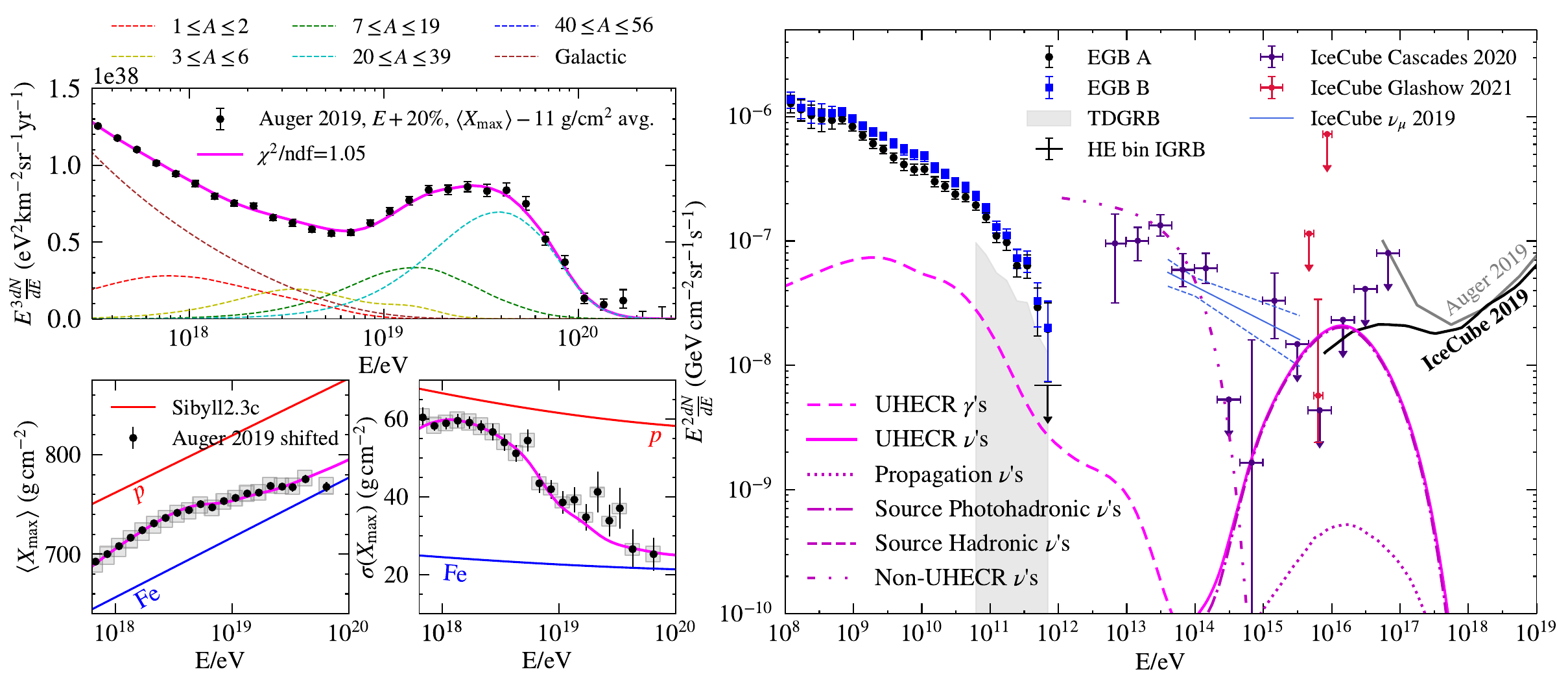}
	\caption{Figure 1 from~\cite{mfu22}, extending the analysis of~\cite{ufa15}. \textbf{Left:} The CR predictions for spectrum (top) and composition (bottom) compared to Auger observations. \textbf{Right:} The neutrino and gamma-ray predictions for this model (solid and dashed lines, respectively). The total neutrino flux due to UHECRs (``UHECR $\nu$'s'', solid magenta) is broken down by origin: UHECR interactions during extragalactic propagation (``Propagation $\nu$'s'', dotted dark magenta), UHECR photohadronic and hadronic interactions in the source (``Source $\nu$'s'', dot-dashed and dashed dark magenta, respectively). Neutrinos originating from a source other than UHECRs (``Non-UHECR $\nu$'s'', dot-dot-dashed dark magenta) are also shown. The observed and inferred values of the extragalactic gamma-ray flux, astrophysical neutrino fluxes, flux measurements from the Glashow event, and upper-bounds on the EHE cosmic neutrino flux from IceCube (black) and Auger (grey) are shown. See~\cite{mfu22} for details.}
	\label{fig:astroNuBestFits_sibyll}
\end{figure*}

As noted in the Introduction, the spectrum and composition of extragalactic UHECRs below 10 EeV is well-explained by the ``UFA'' mechanism~\cite{ufa15}, named for its authors Unger, Farrar and Anchordoqui.  The UFA mechanism abstracts general features of the UHECR accelerator and its surroundings, independently of how they are realized in any particular physical system; the aim was to constrain properties of the sources as a guide to excluding or favoring particular source candidates and pinpointing the locus of the accelerator within the astrophysical system.  Properties of the environment leave their mark on the detailed shape of the spectrum and composition observables as a function of energy, making it possible based on UHECR data to constrain certain features of the environment such as the temperature and magnetic confinement of the surroundings~\cite{mf23}. A central idea of the UFA mechanism is that the light, extragalactic UHECRs below the ankle are produced by spallation of higher energy nuclei in the environment surrounding the accelerator.  The UFA mechanism explains the UHECR data well, as seen in the left panels of Fig.~\ref{fig:astroNuBestFits_sibyll} reproduced from Ref.~\cite{mfu22}, and is the only known way to explain the light sub-ankle population without postulating a separate source class which accidentally smoothly matches in normalization the heavier, higher energy component.  The analysis predicting the neutrino spectrum shown in the right panel of Fig.~\ref{fig:astroNuBestFits_sibyll} should be repeated using the sech cutoff appropriate for magnetized turbulence, which gives an even better description of the UHECR data~\cite{cfm24}. 

The photo-hadronic interaction of a UHECR nucleus as it interacts with the radiation field surrounding the accelerator, can produce a pion or simply break up the nucleus.  A produced pion typically carries about 10\% of the initial nucleon energy;  about half the pions are charged and decay sequentially through the weak interactions to 3 neutrinos, each carrying on average 1/4 of the pion energy.  For, e.g., 20 EeV silicon, which is a dominant nucleus in this energy range, the neutrinos have a typical energy of 20/28/10/4 EeV $\approx$ 18 PeV.   The right panel of Fig.~\ref{fig:astroNuBestFits_sibyll} shows the predicted $\nu$ spectrum based on a more detailed analysis in the UFA framework ~\cite{mfu22}, with parameters chosen to maximize neutrino production consistent with upper limits.  (Note that with the greater specificity of the BNS merger scenario, the UFA approximation can be refined and predictions may vary at the factor-few level.)  Three neutrinos have been reported in this energy range and given that the predictions are at the edge of the limits, it would be natural for more UHECR-produced EHE neutrinos to be observed in the not-distant future;  see~\cite{muf19,mfu22}.  

\noindent  \uline{Fluence of EHE neutrinos accompanying GW170817:}
\hspace{-0.05in}
Here, we place an upper limit on the fluence of neutrinos expected in association with GW170817, to check its consistency with the multimessenger limits. ANTARES, IceCube, Auger and LIGO~\cite{Auger+GWNeutrinos17} placed a limit of about $2.5 {\rm \,GeV cm}^{-2}$ on the integrated fluence of neutrinos accompanying GW170817 in the decade 10-100 PeV. As we shall see, the expected neutrino fluence from GW170817 associated with its UHECR production is far below this sensitivity.  To predict the actual neutrino production per UHECR above 10 EeV, the analyses of~\cite{mfu22} needs to be repeated, recording that information;  I thank M. Muzio for suggesting the shortcut approach used below.  

From Fig.~3 in~\cite{mfu22}, one can see that for Si above 10 EeV, the escape length from the accelerator is 10 to 100 times greater than the photohadronic interaction length.  Only a fraction of the interactions produce a pion, but to derive a conservative bound let us assume 100 pion-producing interactions for each UHECR produced above 10 EeV. Thus the total number of UHECR-produced neutrinos in the 20 PeV peak, per merger, has the upper limit
\begin{equation}
\label{eq:nuperBNS}
\begin{split}
   N_\nu & <  150 \,N_{UCR}  = 150 \, \dot{\mathcal{Q}}/\!<\!E_{UCR}\!> 
    \approx 3/\Gamma_3 \times 10^{39},
\end{split}
\end{equation}
where $N_{UCR}$ is the number of UHECRs per merger, and in the last line we estimated the average energy of CRs above 10 EeV as being 20 EeV.   

The GW170817 merger occurred at a distance of $40.7 \pm 1.4 \pm 1.9$ Mpc~\cite{Cantiello+dist170817}, so with an average energy of $\approx 20$ PeV per neutrino the maximum fluence of $\nu$'s in the 10-100 PeV range is predicted to be less than $3/(\Omega\Gamma_3) \times 10^{-7}\, {\rm GeV \,cm^{-2}}$, where $\Omega$ is the fraction of solid angle of $\nu$ emission, approximately (1-cos $\theta)= 0.02$ for production in a 12$^\circ$ opening angle.  The upper limit on the fluence, even at the peak of a perfectly aligned jet of neutrinos and with the lowest BNS merger rate, $\Gamma_3 = 0.01$, is $\approx 10^{-3}\, {\rm GeV \,cm^{-2}}$; this is far below the upper bound shown in Fig.~2 of ~\cite{Auger+GWNeutrinos17}: $E^2 F\lesssim 2.5 \,{\rm GeV\, cm^{-2}}$.  
Thus detection of a high energy neutrino coincident with GW170817 is not expected, even under the most favorable circumstance of production in a jet aiming directly at Earth.  Therefore the non-detection of an EHE neutrino correlated with GW170817 is not informative as to the validity of the proposed BNS origin of UHECRs.  

\noindent  \uline{Correlations of neutrinos above 5 PeV with short GRBs:}
\hspace{-0.05in}
 Some fraction of GWs have associated short GRBs.  Here we ask if any of the highest energy neutrinos detected so far are correlated with a detected sGRB. 
The three highest energy neutrinos reported to date have central energy estimates of 6.3~\cite{IceCubeGlashow21}, 8.7~\cite{IceCubeNuMu16}, and 11.9 PeV~\cite{Krauss+IC19_20}.  The first is IceCube's ``Glashow event'', which being a cascade event is poorly localized so not useful for a correlation search.  The other two have no counterpart candidates among Swift short GRBs.  

However observing an EHE-neutrino-sGRB correlation in today's data is \textit{a priori} unlikely, because only a fraction of BNS mergers produce sGRBs which have been detected.   An upper limit on the fraction of BNS mergers within $z \leq 1$ producing sGRBs which have been detected and well-localized, $f_{\rm sGRB}$, can be roughly estimated as follows.  With a BNS merger rate of $\Gamma_3 10^3 \,{\rm Gpc}^{-3} {\rm yr}^{-1}$  (see discussion in the main text) and a co-moving volume within $z=1$ of about 160 Gpc$^3$~\cite{wrightCosmoCalc06}, there have been $\approx 2.6 \times 10^6 \,  \Gamma_3 $ BNS mergers within $z\leq 1$, between 2005-2021.  Fong et al.~\cite{Fong+sGRB22} develop a sample of 84 sGRBs with host associations during this time-period, comprising  $\approx$ 56\% of all Swift short GRBs detected from 2005-2021; about 3/4 of these originate within $z \leq 1$.  Therefore an upper limit on the fraction of BNS mergers within $z=1$ accompanied by detected sGRBs is $f_{\rm sGRB} \approx 3 \times 10^{-5}$.  Hence sGRBs are not an efficient proxy for GWs from neutron star mergers.

\noindent  \uline{Future prospects for correlations of neutrinos above 5 PeV with BNS mergers:}
\hspace{-0.05in}
Establishing coincidences between UHECR-produced neutrinos and BNS mergers detected with GW detectors may be feasible with next-generation neutrino and GW detectors. From Fig.~24 of the IceCubeGen-2 Technical Design Report~\cite{ICTDR}, one can deduce that the  sensitivity of IceCube-Gen2 to diffuse astrophysical neutrinos after 10 years of operation should be an order of magnitude below the 20-PeV-peak flux predicted in~\cite{mfu22}, suggesting that of order 10 or more such neutrinos could be detected in 10 years of running.  In the scenario proposed here, every one of those neutrinos has an associated GW produced by the same BNS merger which should arrive moments to years before.  According to Ref.~\cite{GWforecast23}, the efficiency for detection of a BNS merger within $z=2$ at SNR $\geq 10$ should be 40\% with the ``CE4020ET" GW detection network.  If the directional localization of the GW source is good and the time separation is not too large, coincident observations of a BNS merger and an EHE neutrino, which would be a smoking gun for the BNS origin of UHECRs, looks promising with next-generation detectors.   The fraction of BNS mergers identified through their GW signal, in which neutrinos are and are not detected, in conjunction with the known sensitivity for neutrino detection, would test the UHECR-origin hypothesis. Thanks to the rarity of EHE neutrinos, those in the UHECR-produced 20 PeV peak, accidental coincidences should be negligible unless the time separation passes some limit which needs to be quantified.  Given an adequate sample of coincident GWs and neutrinos in the 20 PeV peak, the distribution of time delays between GW and neutrino would provide valuable information on the UHECR acceleration process and conditions in the source environment.   

\section{3. Amaterasu and Andromeda}\label{sec:SMAndromeda}  

The source locus of Telescope Array's Amaterasu event is devoid of distinctive source candidates~\cite{ufAma23}, but does contain our neighbor Andromeda (M31), a galaxy of comparable mass to the Milky Way at a distance of 765 kpc.  In a scenario such as ours where every galaxy hosts UHECR production in proportion to its BNS merger rate, M31 should be an {\it a priori} relatively powerful source.  This raises the question of whether -- were M31 the source of Amaterasu -- could that somehow be discovered and relic evidence of the event investigated.  

To gauge the likelihood that a BNS merger in M31 could have produced Amaterasu, we calculate the expected number of BNS mergers in M31 contributing to UHECRs detected today.  Due to our poor understanding of the intervening magnetic fields, the arrival time delay of a UHECR from M31 is very uncertain. Analysis of Milky Way double-NS systems leads to an estimated BNS merger rate in the Milky Way of $32^{+19}_{-9} \, {\rm Myr}^{-1}$~\cite{DNSmerger_grunthal+21}, which should be applicable to M31 given its similarity.  Two additional factors enter the probability of observing a UHECR today from a BNS merger in M31 some time in the past:  the arrival-time spread, $\tau$, and the fraction of UHECRs reaching the Milky Way, $f_{\rm contrib}$, discussed below Eq.~\eqref{eq:taueff}.  Both depend on the intervening magnetic fields through the smearing factor $\beta$.
 
 The strength and coherence length of the random magnetic field within the Local Group has not been directly constrained.  Using stacked observations of filaments Ref.~\cite{Carretti+23} estimates the average random field strength at $z=0$ to be $B_{\rm rand}\approx (8-26) \, {\rm nG} (\delta_g/10)^{2/3} $, where $\delta_g$ is the gas density relative to the cosmic mean gas density.  But it is not clear what to take for the path-averaged value of $\delta_g$.  Amaterasu's trajectory would likely have passed through the stellar disks of both M31 and the Milky Way, where the random field strength reaches $\mu$G levels, but only over a length scale less than 10 kpc. Writing $\beta_{\mathrm M31}= 6\, (B_\mathrm{LG} / 30 \,\mathrm{nG})\,\sqrt{L_{c, \mathrm{LG}} / 40\,\mathrm{kpc}}$, gives an estimated arrival time spread $\tau_{\mathrm M31} = 184 \,\mathrm{kyr} \,(\beta_\mathrm{M31}/6)^2$ and $f_{\rm contrib} \approx 1$, implying there is a reasonable chance that a BNS merger in M31 occurred recently enough to have produced Amaterasu.  
 
 Someday it might become possible to identify sites of BNS mergers through localized over-densities of $r$-process elements or other features and -- very futuristically -- conceivably determine the orientation of the jet from the geometry of $r$-process enhancement.  In a most optimistic world in which magnetic deflections en route from M31 are small enough for the event to have reached Earth with little angular dispersion, the UHECR arrival direction relative to the jet direction  might enable the geometry of UHECR production to be constrained. 
 Otherwise, there are no features currently apparent that would argue that a BNS merger relic was associated with production of Amaterasu. (Note that deflection in the Milky Way has been accounted for in the discussion of~\cite{ufAma23}.)

\end{document}